\documentclass{elsarticle}
\usepackage[utf8]{inputenc}
\usepackage[version=3]{mhchem}
\usepackage{xcolor}
\usepackage{graphicx}

\begin{document}

\begin{frontmatter}
\title{Quantum Dynamics of Photoactive Transition Metal Complexes. A Case Study of Model Reduction}
\author[1,2]{Olga S. Bokareva}
\author[3]{Oliver K\"uhn \corref{cor1}}
\ead{oliver.kuehn@uni-rostock.de}
\cortext[cor1]{Corresponding author}
\affiliation[1]{Leibniz Institute for Catalysis (LIKAT), Albert-Einstein-Str. 29A, 18059 Rostock, Germany}
\affiliation[2]{Institute of Physics, University of Kassel, Heinrich-Plett-Str. 40, 34132 Kassel, Germany}
\affiliation[3]{Institute of Physics, University of Rostock, Albert-Einstein-Str. 23-24, D-18059 Rostock, Germany}

\begin{abstract}
Transition metal complexes for photochemical applications often feature a high density of electron-vibrational states characterized by nonadiabatic and spin-orbit couplings. Overall, the dynamics after photoexcitation is shaped by rapid transitions between states of different character and multiplicity. Even though transient absorption experiments enable characterization in terms of kinetic rates, the complexity of the systems usually prevents a more detailed analysis. Quantum dynamics simulations using quantum chemically determined model Hamiltonians may provide such details. In particular, one is tempted to pursue a model reduction, such as to identify couplings or vibrational modes most relevant for the dynamics. Here, we address how such an endeavor is challenged by the particular nature of transition metal complexes. For that purpose, we performed quantum dynamics simulations for a recently studied  iron(II) homoleptic complex.
\end{abstract}

\end{frontmatter}

\section{Introduction}
Transition metal (TM) complexes are important for a broad range of catalytical, medical, biological, and material science applications such as photocatalysis, molecular memory, cell imaging, etc., see e.g., Refs.~\cite{Esswein_CR_2007, Blankenship_S_2011, Jiao_N_2022, Halcrow2013}. 
The typical size of many natural and artificial photocatalysts and their high density of electronic and vibrational states, together with spin-orbit, relativistic, and environmental effects pose a challenge to quantum chemical and quantum dynamical simulations \cite{Gonzalez_C_2012, Penfold_CR_2018,daniel20_43, Wagenknecht_CCR_2011, Zobel_JA_2021}.
Therefore, most of the reported simulations of TM excited state dynamics employ the trajectory surface hopping (TSH) method, see. e.g., Refs.~\cite{tully90_1061,mai18_e1370, gomez19_8321, papai22_1329, zobel21_3760, Plasser_JCTC_2019}. 
This quasi-classical approach enjoys popularity due to its clear concept and the possibility of studying complex systems with numerous electronic and vibrational degrees of freedom (DOFs).
Of course, one might ask whether all states and modes are equally important for the dynamics.
The answer to this question often will provide a better mechanistic understanding in addition to a means for model reduction towards a setup more feasible for quantum dynamics.
Methods for identifying essential nuclear motions are well established in the field of molecular dynamics \cite{Virshup_JCP_2012, Zauleck_JCTC_2016, Capano_PCCP_2017, Li_JCTC_2017, Hare_CS_2019}. They all rely solely on the distribution of nuclear geometries,
including minimum energy seams of crossing and minimum energy conical intersections \cite{Harabuchi_JCTC_2016}.
However, as pointed out by Mai and Gonz\'alez \cite{mai19_244115}, displacements alone are not sufficient to judge on the importance of a particular mode, as the impact of the amplitude of a given nuclear motion on the wave packet evolution may be very complex.
Instead, information about electronic energy gaps and state couplings has to be taken into account. 
A workflow for model reduction starting from a TSH  simulation has been proposed by G\'omez et al. \cite{gomez19_8321}, which additionally included feedback on the reduced dimensional quantum simulation, such as to improve TSH parameters. 
Trajectory surface hopping was also used by other authors to identify relevant modes for quantum dynamics simulations of TM compounds \cite{papai22_1329}.

As  excited dynamics takes place on the potential energy surfaces (PESs) precomputed with the  chosen quantum-chemical approach or on-the-fly, the outcomes of such computations may strongly vary with the applied computational scheme.
An exemplary homoleptic iron(II) complex $\mathbf{FePS}$ with two push-pull ligands \cite{FePS_exp} has been recently studied by us using the combination of TSH excited states dynamics with optimally-tuned long-range separated functionals \cite{zobel23_1491}.
Although the concept of optimal tuning was introduced to reliably describe charge-transfer properties such as metal-to-ligand charge transfer (MLCT) excitations, the appearance of multiple low-lying metal-centered states makes the choice of an appropriate range-separation scheme more complicated.
Being formally local, those MC states may suffer from uncompensated self-interaction error and, thus, require special attention beyond the classical optimal tuning.
Following this, additional criteria may be desired to adjust the computational scheme.
Assisted by comparison with the experimental absorption spectra and CASPT2 calculations, two suitable sets of range-separation parameters were chosen for excited state surface hopping dynamics simulations.
The suggested dynamics mechanisms were notably different for both sets of parameters.
Using the set of optimal parameters from the classical $\Delta$SCF method (set~A) resulted in the population of \ce{^3MLCT} states that remained stable within the 2~ps simulation time, which is at variance with transient absorption experiments  \cite{FePS_exp}.
Applying  multi-reference CASPT2-derived parameter set, TSH calculations led to relaxation processes into hot singlets states. 
In the course of the deactivation, the system stayed in regions of high density of states, reachable for further transfer into quintet states in accordance with experimental findings \cite{FePS_exp}.

In what follows, we will use the CASPT2-based optimized functional to illustrate some aspects of quantum dynamics simulations of TM compounds alike $\mathbf{FePS}$. It should be stressed that the purpose is \textit{not} to obtain converged results for the total system, although, in principle, this should be feasible. Instead, we will highlight some issue which one may encounter when trying to design a reduced model for a system with a large density of states and a large number of couplings of comparable magnitude. In what follows, we first outline the linear vibronic coupling (LVC) model used for the quantum dynamics simulations. Next, we address the Multi-layer Multiconfiguration Time-dependent Hartree (ML-MCTDH) method. In the applications section, we first characterize the LVC model for the $\mathbf{FePS}$ system. Possible model reduction is investigated using the Time-dependent Hartree (TDH) limit of MCTDH. For the thus identified reduced model, ML-MCTDH simulations are performed to address correlation effects. Finally, conclusions are given. 

\section{Theory}
\subsection{Linear Vibronic Coupling}
In contrast to trajectory-based methods,  quantum dynamics simulations require a priori knowledge of the PES, including possible electronic state couplings. For TM complexes, this inevitably means that one has to resort to model PES, the most prominent one being the vibronic coupling model \cite{koppel84_59,penfold18_6975,may23_}. It rests on a harmonic oscillator description of nuclear motion in terms of normal modes, combined with a low-order expansion for the displacement of the PESs and the state couplings. Specifically, the linear  LVC Hamiltonian reads, denoting the set of $f$ normal mode coordinates as $\mathbf{Q}$,
\begin{equation}
    H=\sum_{mn}[\delta_{mn}H_m(\mathbf{Q})+  V_{mn} (\mathbf{Q})] |m\rangle\langle n|
    \label{eq:ham}
\end{equation}
with the Hamiltonian for electronic state~$m$ 
\begin{equation}
    H_m(\mathbf{Q}) = E_m+\frac{1}{2}\sum_\xi \hbar\omega_\xi (-\partial^2_\xi +Q_\xi^2) +\sum_\xi \kappa_{m,\xi}Q_\xi
\end{equation}
and the interstate couplings
\begin{equation}
    V_{mn} (\mathbf{Q}) =V_{mn}^{(0)}+\sum_\xi \lambda_{mn,\xi}Q_\xi\,.
    \label{eq:vmn}
\end{equation}
Here, $\kappa_{m,\xi}$ and $\lambda_{mn,\xi}$ are the intra- and interstate coupling constants, respectively, related to mode $Q_\xi$. Further, $V_{mn}^{(0)}=V_{mn}^{(\rm SOC)}$ denotes the spin-orbital coupling (SOC), assuming that its coordinate dependence is negligible. The Hamiltonian, Eqs.~(\ref{eq:ham} - \ref{eq:vmn}), can be supplemented by 
\begin{equation}
    H_{\rm f}(t) = - {\mathbf E}(t) \sum_{mn} {\mathbf d}_{mn} |m\rangle\langle n|
    \label{eq:hfield}
\end{equation}
describing the interaction with an external time-dependent electric field ${\mathbf E}(t)$ in dipole approximation (${\mathbf d}_{mn}$ is the vector of the dipole matrix elements).

Note that the LVC Hamiltonian is not only used for quantum dynamics simulations (see, e.g. Refs. \cite{Fumanal_JPCL_2018, Papai_JCP_2019}), if applicable, it also provides an ideal framework for TSH. Fortunately, this often holds true for TM complexes where on-the-fly generation of PES is computationally very expensive \cite{zobel21_3760,zobel23_1491}.
This model is only valid if no large amplitude motions in the excited state dynamics are present, for diagnostics of accuracy, see the exemplary study in Ref.~\cite{Penfold_PCCP_2023}.
In cases where bond breaking is simulated, the LVC model for MCTDH may be complemented by the corresponding dissociative mode, see e.g. studies of CO-release from TM-complexes \cite{Falahati_NC_2018, Fumanal_JCP_2021}.

\subsection{ML-MCTDH Method}
The method of choice for high-dimensional quantum dynamics simulations of LVC Hamiltonians is the MCTDH approach \cite{meyer90_73,beck00_1,wang03_1289,meyer11_351,vendrell11_044135}, which is briefly sketched in what follows. Given the expansion of the state vector in the diabatic electronic basis
\begin{equation}
|\Psi({\bf Q};t) \rangle=\sum_{m} \psi_{m}({\bf Q};t) \, |m\rangle	\, .
  \label{eq:psi_total}    
\end{equation}
The state-specific vibrational wave packets are represented as follows (skipping the state index $m$ for simplicity)
\begin{equation}
    \psi({\bf Q};t)=\sum_{j_1=1}^{n_1}\ldots \sum_{j_f=1}^{n_f} A_{j_1 \ldots j_f}^1(t)\cdot\phi_{j_1}^{1;1}(Q_1,t)\ldots \phi_{j_f}^{1;f}(Q_f,t)\, .\, .
\label{eq:mcpsi}
\end{equation}
Here, $\phi_\lambda^{1,\kappa}(Q_\kappa,t)$ are the so-called single-particle functions (SPFs) which form a time-dependent basis for expanding the wave packet in terms of different Hartree products of SPFs. Eq.~(\ref{eq:mcpsi}) provides a compact representation such that the number of SPFs per DOF, $n_\kappa$, usually will be well below that required for a static basis. This ansatz can, in principle, be converged towards the numerically exact limit. On the other hand side, there  is the  TDH limit, where only a single SPF per DOF is taken into account. In practice, the SPFs are expanded into a time-independent (primitive) basis set,  $\chi_j^{(\kappa)}(Q_\kappa)$, according to 
\begin{equation}
 \phi^{1;\kappa}_\lambda (Q_\kappa,t)=\sum^{N_\kappa}_{j=1}A^{2;\kappa}_{\lambda;j}(t) \chi_j^{(\kappa)}(Q_\kappa) \, .
 \label{eq:chi}
\end{equation}
MCTDH can be seen as a two-layer approach.  This view becomes clear upon interpretation of $A^{2;\kappa}_{\lambda;j}(t)$ as an additional set of expansion coefficients to represent the time-dependent SPFs of the upper (first) layer in a time-independent primitive basis in the lower (second) layer. In $A^{2;\kappa}_{\lambda;j}(t)$ the superscripts 2 and $\kappa$ correspond to the second layer and the $\kappa$th mode, respectively, whereas the subscript $\lambda$ refers to the indices of the respective SPF of the  first layer.  The expansion coefficients in Eqs.~(\ref{eq:mcpsi}) and (\ref{eq:chi}) can be obtained using the time-dependent Dirac-Frenkel variational principle \cite{beck00_1}.

ML-MCTDH builds on the idea of combining strongly correlated DOFs into $p$ logical coordinates  (combined modes or MCTDH particles) yielding  the  set $\{q^1_1,\ldots,q^1_p\}$ with $q^1_\kappa= \{Q_{1_\kappa},\ldots,Q_{d_\kappa}\}$, where the superscript refers to the first layer, and $d_\kappa$ is the dimension of the $\kappa$th particle. The wave packet within the combined modes' picture for the uppermost (first) layer can be written as
\begin{align}
\Psi(q^1_1,\ldots,q^1_p,t)=\sum_{j_1=1}^{n_1}\ldots\sum_{j_p=1}^{n_p} A_{j_1\ldots j_p}^1(t)\,\phi_{j_1}^{1;1}(q^1_1,t)\ldots\phi_{j_p}^{1;p}(q^1_p,t)\, .
\label{psicomb}
\end{align}
In ML-MCTDH, the multi-dimensional SPFs for the particles, $\phi_{\lambda}^{1;\kappa}(q^1_\kappa,t)$, are considered as multi-dimensional wave packets of the second layer and expanded according to
\begin{eqnarray}
\phi_{\lambda}^{1;\kappa}(q^1_\kappa,t)&=&\phi_\lambda^{2;\kappa}(Q_1^{2;\kappa},\ldots,Q_{d_\kappa}^{2;\kappa},t)\nonumber \\
&=& \sum\limits_{j_1}^{n_{\kappa,1}}\ldots\sum\limits_{j_{d_\kappa}}^{n_{\kappa,d_{\kappa}}}A^{2;\kappa}_{\lambda;j_1\ldots j_{d_\kappa}}(t) \,\phi_{j_1}^{2;\kappa;1}(Q_1^{2;\kappa},t)
\ldots \phi_{j_{d_\kappa}}^{2;\kappa;d_\kappa}(Q_{d_\kappa}^{2;\kappa},t) \,.
\label{eq:mlmctdh1}
\end{eqnarray}
In the third layer, these wave packets are represented in the primitive basis, i.e.
\begin{align}
\phi_{\lambda}^{2;\kappa;\sigma}(Q_{\sigma}^{2;\kappa},t)=\sum\limits_{j=1}^{N_\alpha}A^{3;\kappa;\sigma}_{\lambda;j}(t)\, \chi_j^{(\alpha)}(Q_{\sigma}^{2;\kappa}), \quad \quad \alpha=\sigma+\sum\limits_{i}^{\kappa-1}d_i \,.
\label{eq:mlmctdh2}
\end{align}
While this example included three layers, the general idea can easily be applied to a case with more layers containing particles with decreasing dimensionality when stepping towards the bottom layer.
Within this scheme, electronic states are incorporated by introducing an electronic coordinate, $Q_{\rm el}$, in the uppermost layer. 
By construction, ML-MCTDH allows for a numerical solution of the time-dependent Schr\"odinger equation with controlled convergence. 
Finally we note that the effect of an external laser field, Eq.~(\ref{eq:hfield}), can be incorporated into the MCTDH scheme  a straightforward manner \cite{naundorf02_719}. For the simulations present below the Heidelberg program package was used \cite{mctdh86}. 
%
\section{Application}
\subsection{LVC Hamiltonian for $\mathbf{FePS}$}
In what follows, we will apply MCTDH to study the quantum dynamics of an exemplary iron(II) homoleptic complex [Fe(cpmp)$_{2}$]$^{2+}$, where cpmp = 6,2’’-carboxypyridyl-2,2’-methylamine-pyridyl-pyridine, shown as an inset in Figure~\ref{fig:spec} and abbreviated as $\mathbf{FePS}$  \cite{FePS_exp}. 
The choice is motivated by our previous investigation of the influence of the parametrization of the range-separated hybrid  functional on the photoinduced dynamics as described by TSH. 
However, the present focus will not be on a quantitative comparison with the TSH dynamics reported in Ref.~\cite{zobel23_1491}, but on coping with the high dimensionality of the LVC Hamiltonian for $\mathbf{FePS}$ in general. As an example, we use the 
parameter set of the functional derived based on a comparison with CASPT2 results, which gave a better agreement between calculated and experimental dynamics. In numbers, taking into account the lowest 9~singlet and 14~triplet states (energies up to about 2.8~eV), there are in total 51~electronic states. Further, the molecule has 212~vibrational DOFs.

\begin{figure}[tbh]
    \centering
    \includegraphics[width=1\textwidth]{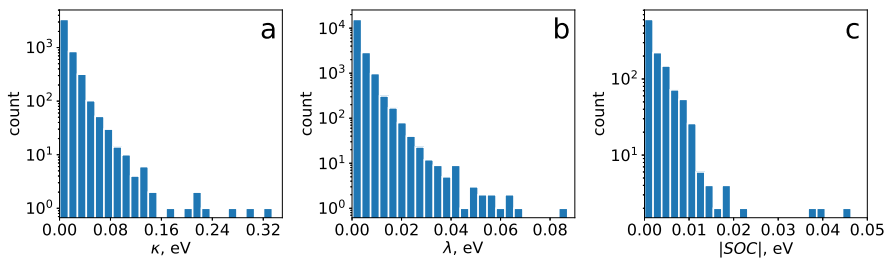}
    \caption{Distribution of LVC and SOC coupling parameters: (a) $\kappa=|\kappa_{m,\xi}|$, (b) $\lambda =|\lambda_{mn,\xi}|$, and (c) $|{\rm SOC}| = |V_{mn}^{(\rm SOC)}|$.}
    \label{fig:lcvdist}
\end{figure}

\begin{figure}[tbh]
    \centering
    \includegraphics[width=0.6\textwidth]{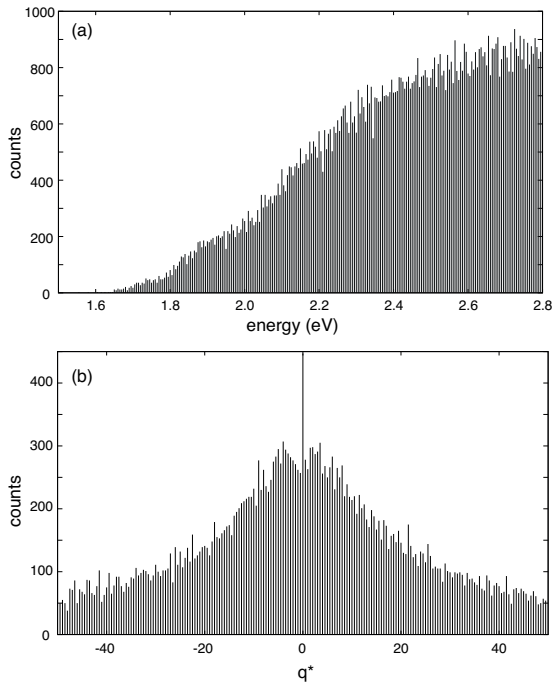}
    \caption{Analysis of the LVC model for  $\mathbf{FePS}$ \cite{zobel23_1491}. (a) Distribution of diabatic electron-vibrational spin-free singlet and triplet states (bin width 5 meV). (b) Distribution of state crossing for all possible transitions between spin-free singlet and triplet states (bin width 0.5).}
    \label{fig:dos-qstar}
\end{figure}

Let us start with an analysis of the LVC Hamiltonian reported in Ref. \cite{zobel23_1491}. Assuming an energy threshold of 0.0003~eV, the LVC Hamiltonian contains 4722~$\kappa_{m,\xi}$ terms, 20119~$\lambda_{mn,\xi}$ terms, and 2464~SOC terms. Figure~\ref{fig:lcvdist} shows the distribution of LVC parameters, which indicates that the majority of couplings is rather small. In Figure~\ref{fig:dos-qstar}a, we have plotted the distribution of zeroth-order states, defined as spin-free states without nonadiabatic state coupling ($\lambda_{mn,\xi}=0$). Apparently, the number of states increases rather rapidly with energy.

Since it is known that TM  compounds often feature a rather rapid  ISC dynamics, it is interesting to inspect the crossing points between singlet and triplet states. According to the LVC Hamiltonian, for a pair of zeroth-order states, they are given by $q_\xi^*=(E_n-E_m)/(\kappa_{m,\xi}-\kappa_{n,\xi})$. The distribution of crossing points for $\mathbf{FePS}$ is shown in Figure~\ref{fig:dos-qstar}b. To put these numbers into perspective, one should note that the mode elongation at the zero point energy ranges between $\pm$1 (in dimensionless units). The actual values for $\mathbf{FePS}$ are mostly much larger. The reason is to be found in the rather small displacements, cf. Figure~\ref{fig:lcvdist}. This observation is no contradiction to a  rapid ISC since the potential curves run almost parallel until they eventually cross. Since SOC is assumed to be not coordinate dependent, this facilitates an efficient ISC if energies of spin-free states are energetically not too far apart. 

\subsection{TDH Dynamics}
\label{sec:tdh}
Obtaining converged MCTDH dynamics for the given model on a few picosecond time scale should be possible with the multilayer approach but is not the focus of this contribution. 
Instead, we will use the TDH limit of the MCTDH ansatz, which allows for full-dimensional quantum simulations \cite{paramonov01_205}. 
The TDH limit facilitates the  exploration of  overall aspects of the dynamics of different reduced models on equal footing. Note that for ML-MCTDH calculations, each model should be converged separately.  The analysis in the previous section showed that the LVC model is characterized by a high density of states and a large number of couplings terms having a comparable magnitude. In this section, we will address the question of whether this situation allows for a model reduction by defining certain  cut-offs for the coupling parameters. 

\begin{figure}[tbh]
    \centering
    \includegraphics[width=0.9\textwidth]{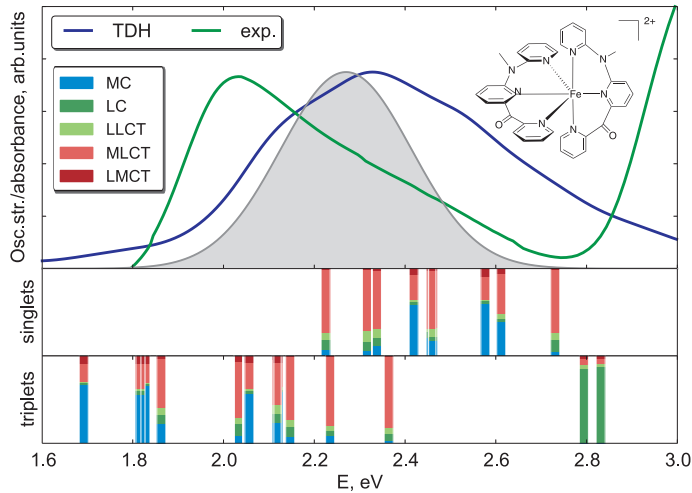} 
    \caption{Upper panel: Absorption spectrum of $\mathbf{FePS}$ as calculated by the TDH approach to the dipole autocorrelation function as well as experimental results \cite{FePS_exp}. Also shown is the laser pulse spectrum (gray shaded area). Lower panels: Decomposition of electronic transitions based on density matrix analysis \cite{Plasser_JCP_2020}.}
    \label{fig:spec}
\end{figure}

Figure \ref{fig:spec} compares the experimental absorption spectrum with the one calculated with the correlation function approach using TDH dynamics of the full model. Further, an assignment of spin-free electronic transitions is given. The overall lineshape of the spectrum is reasonably reproduced; in addition to the inherent broadening due to the large density of states, we have added a phenomenological broadening of $\gamma^{-1}=10$~fs to account for solvation effects. As compared with experiments, there is a shift of about 0.3~eV. Since this range of the experimental spectrum does not convey many details, model reduction will be investigated for laser-driven dynamics.

For the following simulations, we will consider the $x$-component of the dipole operator only (which is of  higher magnitude  than the other components). Further we  assume a Gaussian pulse shape, i.e.
\begin{equation}
E(t)=E_0 \cos(\Omega t) \exp\{-(t-t_0)^2/2\sigma^2 \} \,.    
\label{eq:field}
\end{equation}
Here, $\Omega$, $t_0$, and $\sigma$ are the carrier frequency, the pulse center, and the pulse width, respectively. The transition energy at the peak of absorption due to the $x$-component of the dipole  is at 2.27~eV. For the width parameter, we use 24~fs yielding a full-width a half maximum of the pulse intensity of 40~fs. The field amplitude has been chosen as 0.001~Hartree/Bohr, which gave a population transfer of about 25-30\% to the excited singlet states for the TDH dynamics.

\begin{figure}[tbh]
    \centering
    \includegraphics[width=0.7\textwidth]{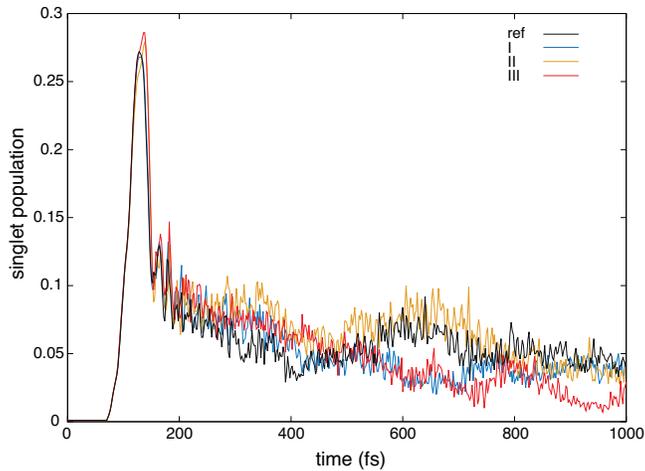}
    \caption{TDH dynamics of summed excited singlet state populations after excitation with a $x$-polarized  Gaussian laser pulse centered at $t_0=100$ fs ($\hbar \Omega=2.27$~eV, $\sigma=24$~fs, $E_0=0.001$~Hartree/Bohr). The different curves correspond to the full model (denoted 'ref') and various reduced modes as explained in the text.}
    \label{fig:popTDH}
\end{figure}

Figure~\ref{fig:popTDH}  compares the population of all singlet states (except the ground state) for different models. The reference case (full model) is characterized by an initial rise due to the laser pulse and an immediate rapid decay. (Note that this decay is not due to stimulated emission.) Subsequently, one observes a slower decay with a partial revival around 600-700~fs as well as superimposed rapid oscillations. The dynamics of the other three models shown in the figure essentially agree during the first 200~fs. Putting the threshold for the nonadiabatic couplings to $|\lambda_{mn,\xi}| = 1$~meV (I, blue line), the decay after 200~fs is a bit slower, and the partial revival doesn't occur. Using, in addition, the threshold $|\kappa_{m,\xi}|=5$~meV (II, orange line) brings back the revival. Putting in addition a threshold for SOC of $|V_{mn}^{(\rm SOC)}|=0.5$~meV (III, red line), there is again no pronounced revival. It should be stressed that these parameters have been taken after consideration of certain larger cut-off values such as $|\lambda_{mn,\xi}| = 5$~meV, which gave a pronounced difference in the dynamics. (Note that for the purpose of illustration, it was not attempted to pinpoint these parameters in more detail.) At first glance, the small magnitude of these cut-off parameters is a bit surprising. In fact, inspecting Figure~\ref{fig:lcvdist} one could have been tempted to select just a few modes having with the largest couplings. Nevertheless, in the present case of $\mathbf{FePS}$, the dynamics after laser excitation appears to be shaped by the combined effect of many small-magnitude couplings. We note that the absorption spectrum is essentially indistinguishable for these four models. The  resulting LVC Hamiltonian with all three coupling parameters reduced contains 2941~$\kappa_{m,\xi}$ terms, 13243~$\lambda_{mn,\xi}$ terms, and 2166~SOC terms.

\begin{figure}[tbh]
    \centering
    \includegraphics[width=0.8\textwidth]{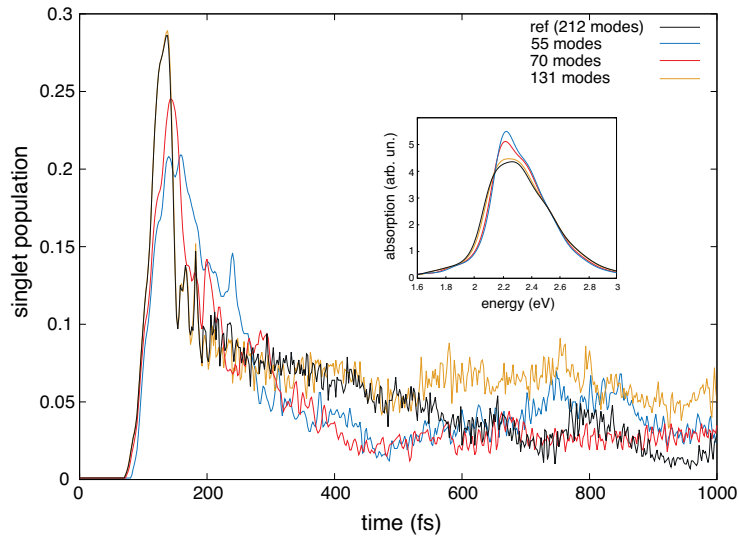}
    \caption{Comparison of the population dynamics for model III of Fig. \ref{fig:popTDH} (212 modes) with various models having a reduced number of vibrational modes as indicated; for selection details see text. The inset shows the absorption spectrum for the four modes according to the $x$-component of the dipole.}
    \label{fig:popmodesTDH}
\end{figure}

Next, we explore a possible reduction of the number of vibrational DOFs. Intuitively one would be tempted to compare the coordinate expectation values with the values of the crossing points $q_\xi^*$ between singlet and triplet states. However, according to Figure~\ref{fig:dos-qstar}b, the majority of  $q_\xi^*$ takes rather large values which are not reached during the dynamics. As a technical note, we would like to mention that for the simulation, the single-set method has been used where a given SPF is optimized for all electronic states and coordinate expectation values are calculated accordingly. Figure~\ref{fig:popmodesTDH} compares the reduced model identified above with models having different cut-offs for the average normal mode elongation. Namely, the cut-off has been set to 0.5 (24~modes), 0.2 (70~modes), and 0.1 (131~modes). First, we note that in contrast to Figure~\ref{fig:popTDH}, the initial dynamics during the first 200~fs is influenced by the cut-off. In fact, inspecting the absorption spectrum shown in the inset, one notices marked changes upon reduction of the number of modes. The change in cut-off modifies the overlap with the pulse spectrum and, thus, the wave packet created on the excited state manifold. To have reasonably accurate initial excitation and, at the same time, a considerable reduction of the number of modes, we identified the model with 70~modes as a compromise. Again, we emphasize that the thresholds are rather small and not that much different, i.e. it is the net effect of a large number of small-amplitude vibrations, which is responsible for the observed dynamics.

\subsection{ML-MCTDH Dynamics}

TDH is the mean-field approximation to MCTDH and, as such, neglects correlations in the dynamics of the different DOFs. One could argue that in cases of a large number of small couplings combined with a large density of states, the mean-field approximation should work well. However, since $\mathbf{FePS}$ already gave some surprises, it will be better to compare TDH with MCTDH dynamics, in particular, since ML-MCTDH provides a means to obtain numerically converged results. For the purpose of illustration, we will use the final 70~ modes model obtained from the TDH simulations in the previous section. ML-MCTDH requires to identify a ML-tree to structure the wave packet. This procedure is a nontrivial task, and different trees might lead to rather different CPU usage for obtaining the same accuracy \cite{schroter15_1}. Presently, we did not attempt to optimize the tree structure. Instead, we started with two branches, one including all modes with an average elongation beyond 0.5 (21~modes) and one including the remaining 49~modes. The first branch was further subdivided into three branches, each containing two two-dimensional and one three-dimensional particle. The second branch was first subdivided into four subbranches, each of them containing three further branches. On the  lowest level, there were 23~two-dimensional and one three-dimensional particle. The number of SPFs in each layer has been chosen such that the largest population of the least occupied natural orbital was smaller than 2\%.

\begin{figure}[tbh]
    \centering
    \includegraphics[width=1.3\textwidth]{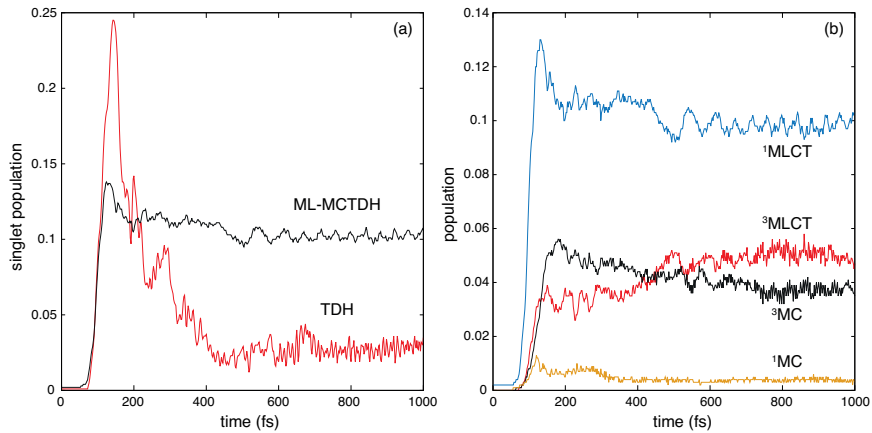}
    \caption{Comparison between TDH and ML-MCTDH dynamics after laser pulse excitation for the reduced model introduced in Section \ref{sec:tdh}. (a) Total population of excited singlet states. (b) Contributions of individual types of states to the ML-MCTDH dynamics as indicated.}
    \label{fig:TDH-MCTDH}
\end{figure}

Figure \ref{fig:TDH-MCTDH}a compares TDH and ML-MCTDH dynamics for the total excited singlet state population. Even though the linear absorption spectrum is essentially identical, the difference in population dynamics  is dramatic. First, ML-MCTDH does not show the pronounced initial peak during and immediately after the laser pulse as is observed for TDH. More notably, however, the decay of the total singlet state population is much slower in the ML-MCTDH case. In fact, this brings the present simulation in closer agreement with the TSH results reported in Ref. \cite{zobel23_1491}. Of course, one might argue that at this point, one should go back to the model reduction performed in Section \ref{sec:tdh} and to  repeat it using ML-MCTDH. However, we don't expect a qualitative difference, i.e., the illustration of the challenges to model reduction performed with TDH will still be valid.

In Figure \ref{fig:TDH-MCTDH}b, ML-MCTDH  populations of individual sets of states are shown. The rapid initial population of triple states is typical for TM compounds and essentially occurs during the laser pulse action. Here, we notice that the most/least population is in the $^1$MLCT/$^1$MC states. The population of 
$^3$MLCT/$^3$MC states is about equal and in between the $^1$MLCT and $^1$MC cases. This general behavior is also in accord with the TSH  results of Ref. \cite{zobel23_1491}. Note that one would not expect a quantitative agreement, not at least because the description of the laser pulse excitation differs in the two simulations (cf. systematic study of this aspect in Ref. \cite{heindl21_144102}).
\section{Conclusions}
We have studied the laser-driven dynamics of the  exemplary \textbf{FePS} complex whose photophysical and quantum chemical properties were previously reported along with TSH simulations. \textbf{FePS} shows the high density of electron-vibrational states typical for such TM complexes. Using an LVC Hamiltonian, the reduction of the number of electronic coupling parameters was explored, followed by a reduction of the number of vibrational modes. The structureless linear absorption spectrum was not useful in this respect, and, therefore, the TDH dynamics was analyzed. It was found that the dynamics is influenced by a large number of small couplings, which prevents the identification of a representative model with just a few DOFs. For the thus established model, still containing 70 vibrational modes and thousands of coupling terms, the effect of correlations on the dynamics was investigated. Here one would have expected that given the nature of the model, effects beyond the mean-field description play a minor role only. Surprisingly, only the inclusion of correlations brings the dynamics in good agreement with previously reported TSH simulations and experiments. 

We expect that the challenges for model reduction discussed here for an exemplary case are typical for similar TM complexes.
\section*{Acknowledgement}
The authors are grateful to Dr.~J.-P.~Zobel for providing the LVC Hamiltonian and L.~Kleindienst for his support with preparing MCTDH input files from LVC data. This work has been financially supported by the Deutsche Forschungsgemeinschaft [Priority Program SPP 2102 ``Light controlled reactivity of metal complexes'' (grant KU952/12-1)].
 \bibliographystyle{elsarticle-num} 
\bibliography{ESKIMO.bib}
\end{document}